\documentclass[fleqn,10pt]{wlscirep}
\usepackage[utf8]{inputenc}
\usepackage[T1]{fontenc}
\usepackage{graphicx}
\usepackage{amsmath}
\usepackage{amssymb}
\usepackage{color}

\begin{document}
\title{
Spread of variants of epidemic disease based on the microscopic 
numerical simulations on networks} 

\author[*]{Yutaka Okabe}
\author[+]{Akira Shudo}
\affil{
Department of Physics, Tokyo Metropolitan University, 
Hachioji, Tokyo 192-0397, Japan}

\affil[*]{okabe@phys.se.tmu.ac.jp}
\affil[+]{shudo@tmu.ac.jp}

\def\l{\langle}
\def\r{\rangle}

\date{\today}

\begin{abstract}
Viruses constantly undergo mutations with genomic changes. 
The propagation of variants of viruses is an interesting problem.  
We perform numerical simulations of the microscopic epidemic model
based on network theory for the spread of variants. 
Assume that a small number of individuals infected with the variant 
are added to widespread infection with the original virus.
When a highly infectious variant that is more transmissible than 
the original lineage is added, the variant spreads quickly to the wide space. 
On the other hand, if the infectivity is about the same as 
that of the original virus, the infection will not spread. 
The rate of spread is not linear 
as a function of the infection strength 
but increases non-linearly. 
This cannot be explained by the compartmental model 
of epidemiology but can be understood 
in terms of the dynamic absorbing state known from the contact process. 
\end{abstract}

\maketitle

\section{Introduction}

Viruses of infectious diseases are constantly changing through mutation 
and become more diverse. It is expected that new variants of a virus 
will arise. Sometimes new variants appear and disappear. 
Other times, new variants may persist. 
It is important to study how viruses change and spread. 
Several novel variants of SARS-CoV-2, the virus that causes COVID-19,
emerged in late 2020 \cite{Davies1,Davies2,Volz}. 
One of these, lineage B.1.1.7, was first detected in the UK 
in September 2020 and spread to multiple countries worldwide. 
This variant was labeled Alpha variant 
by the World Health Organization on 31 May 2021. 
The rapid spread of the UK variant (lineage B.1.1.7) suggests that 
it spreads more efficiently from person to person 
than the existing variants of SARS-CoV-2. 
The South Africa variant (B.1.351, Beta variant) 
and the Brazil variant (P.1, Gamma variant) 
were also identified in many countries. 
More recently, the India variant (B.1.617.2, Delta variant), 
which seems to be highly transmissible, is prevalent in some countries. 
A variant is of concern because it is more likely to spread, 
causing more severe disease, reducing the effectiveness of treatments 
or vaccines, or being more difficult to detect with current tests. 

A variety of epidemiological models have been proposed to analyze 
the spread of the epidemic disease.  Among them, 
a compartmental model is most commonly used, and each individual 
is supposed to be one of the possible types such as susceptible (S), 
infected (I), or recovered (R).  The proportions of individuals 
in each type are taken to be continuous variables, 
and the rate equations among these proportions are then derived. 
The susceptible-infected-recovered (SIR) model, proposed in 1927 
by Kermack and McKendrick~\cite{Kermack}, is the most widely 
recognized basic model belonging to such a class.
In subsection \ref{sub:SIR}of Methods, 
we outline the differential equation of 
the SIR model \cite{Bailey,Diekmann2000,Okabe}, 
because we perform simulations of the microscopic model of epidemic disease, 
which corresponds to the SIR model. 

A given infected individual does not have 
an equal probability of infecting all others. 
Each individual only has contact with a small fraction 
of the total population, and the number of contacts that
people have can vary greatly from one person to another. 
The connection between individuals can be described by a network. 
A network is a set of nodes (vertices) connected by edges. 
Two nodes share one edge, and from the point of view of one node, 
it has a direct relationship with the node connected by the edge, 
which is called a connected (neighboring) node.
The number of edges of a node is referred to as 
the degree of that node. 

Network science is the study of complex networks in the real world. 
It is based on mathematics and has applications in a wide range of fields, 
including statistical physics, computer science, electronics, ecology, 
economics, finance, and public health 
\cite{Caldarelli,Barabasi_book,Newman_book}. 
Here, we briefly describe what is relevant to the complex networks 
used in this study.
The important development in network science was the work 
of Erd\"os and R\'enyi~\cite{Erdos,Erdos2} on graphs 
with random connections between nodes.
This differs from the properties of the graphs actually appearing 
in the real world, but it is well-defined and a good testing ground 
for examining the properties of networks.
Many real-world networks are neither uniform nor random. 
Small-worldness was pointed out; that is, a person can become 
a friend of a friend with only six intermediaries. 
Another point is the existence of a hub; 
there are nodes with a very large number of connections $k$, 
and the distribution of $k$ is 
often scale-free, with a power distribution such as $k^{-a}$ 
\cite{Price,Barabasi}.
Barab\'asi and Albert~\cite{Barabasi} proposed 
an algorithm to create networks that exhibit scale-free properties. 
It is an algorithm that grows networks by selectively combining them. 
This mechanism is called "preferential attachment". 
The emergence of the Barab\'asi-Albert network has led to rapid progress 
in the study of complex networks across many fields.

The importance of the network structure in the analysis of epidemics 
was emphasized, and studies in several directions have been done 
\cite{Pastor,Dezso,Newman,Hufnagel,Keeling,Castellano,Pastor2015}. 
In connection with COVID-19, stochastic simulations of the epidemic 
model have been performed \cite{Herrmann,Choi}.  
The existence of absorbing states is one of the vital non-equilibrium 
processes on the complex network. Once the state has fallen 
into an absorbing state, the dynamics cannot escape 
from it \cite{Marro,Henkel,Mata}. 
The non-equilibrium absorbing phase transition is related to 
the directed percolation \cite{Broadbent}. 
In epidemic spreading processes \cite{Diekmann2000}, a fully healthy state 
can be regarded as an absorbing state in this sense. The contact 
process \cite{Harris} and the susceptible-infected-susceptible 
(SIS) model \cite{Anderson} are often used to describe epidemic dynamics. 
It is well known that the SIS model can be mapped onto 
the logistic equation. 
In the epidemic scenario of the SIS model, 
individuals can be infected or susceptible. 
A phase transition between a disease-free (absorbing) phase and 
an active stationary phase is separated by an epidemic threshold. 
In the latter phase, a fraction of the population is infected.
In the framework of the SIS model, 
the absence of an epidemic threshold 
was discussed for the spreading of infections on scale-free networks 
\cite{Pastor}.

The present authors \cite{Okabe2021} performed simulations 
for the microscopic SIR model on networks, and in particular, 
the relationship between the SIR model for macroscopic quantities 
and the corresponding microscopic SIR model was discussed. 
For the network, we discussed the difference between random networks 
and scale-free networks, and the role of hubs 
in scale-free networks was elucidated. 
The simulation method follows the method of 
Herrmann and Schwartz \cite{Herrmann}. 
They discussed the role of the absorbing state in the SIR model. 

\vspace{2mm}

In this paper, we use a microscopic model of infectious disease 
transmission on networks to simulate the spread of variants. 
The simulation method is an extension of the method used in the absence 
of variants \cite{Okabe2021}.
A related problem is that of "competing epidemics" 
\cite{Newman2005,Karrer,Darabi}.
When there are two competing diseases and the relative infectivity 
of the two diseases is different, a phase diagram 
of the behavior of infection was investigated \cite{Karrer}.

In the following, 
we describe the simulation method of microscopic 
SIR model including variants in subsection \ref{sub:micro} of Methods. 
As for networks, we treat two types of networks;
the Erd\"os-R\'enyi (ER) network \cite{Erdos,Erdos2}, 
a random network, and the Barab\'asi-Albert (BA) network 
\cite{Barabasi}, a scale-free network. 
We will discuss the relationship between the infectious strength 
of the variant and the spread of the infection. 
We emphasize the relation to the non-equilibrium 
absorbing phase transition.

\section{Results}

\subsection{Simulation of the microscopic model 
for the variants on the ER network}

We first consider the case of the ER network, a random network. 
The total number of nodes (individuals) is $N=10000$ and 
the average number of degrees is $\l k \r=8$. 
To perform the simulation of the microscopic model 
for the spread of infection, 
we choose the probability $p$ of infection as $p=9/200$, 
which leads to $\beta = \l k \r p = 0.36$ in terms of 
a rate constant of the SIR model, 
which will be explained in subsection \ref{sub:SIR} of Methods.
The average infected period, $1/\gamma$, is chosen as 
5.0 days, which is realized by a Poisson distribution 
with an average value of 5.0. 
The basic reproduction number 
$R_0=\beta/\gamma$, Equation (\ref{basic_repro}), 
becomes $R_0=1.8$.

\begin{figure}[h]
\centering
{\bf a} \hspace*{7cm} {\bf b}\hspace*{7cm}

\includegraphics[width=7cm]{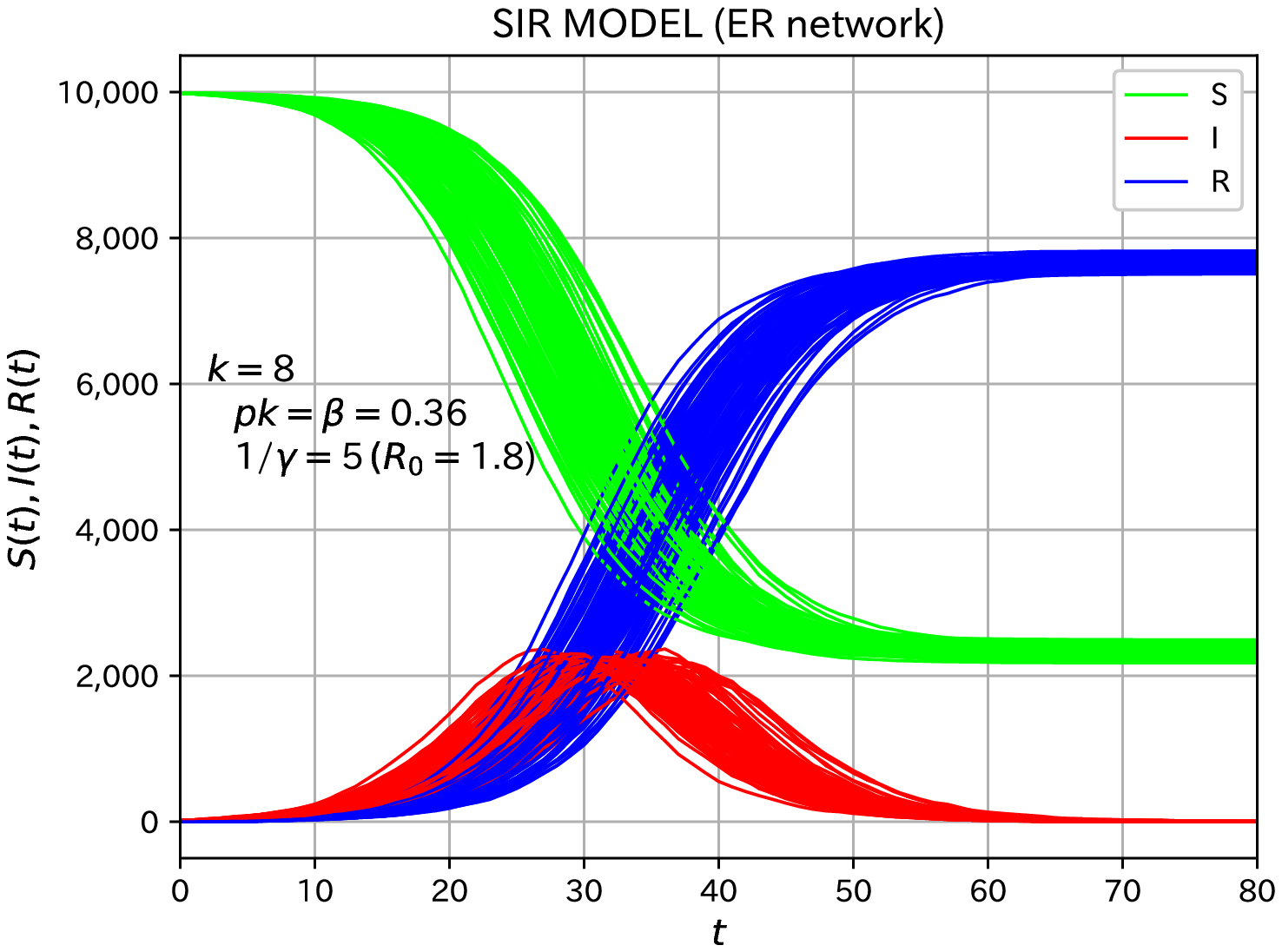}
\includegraphics[width=7cm]{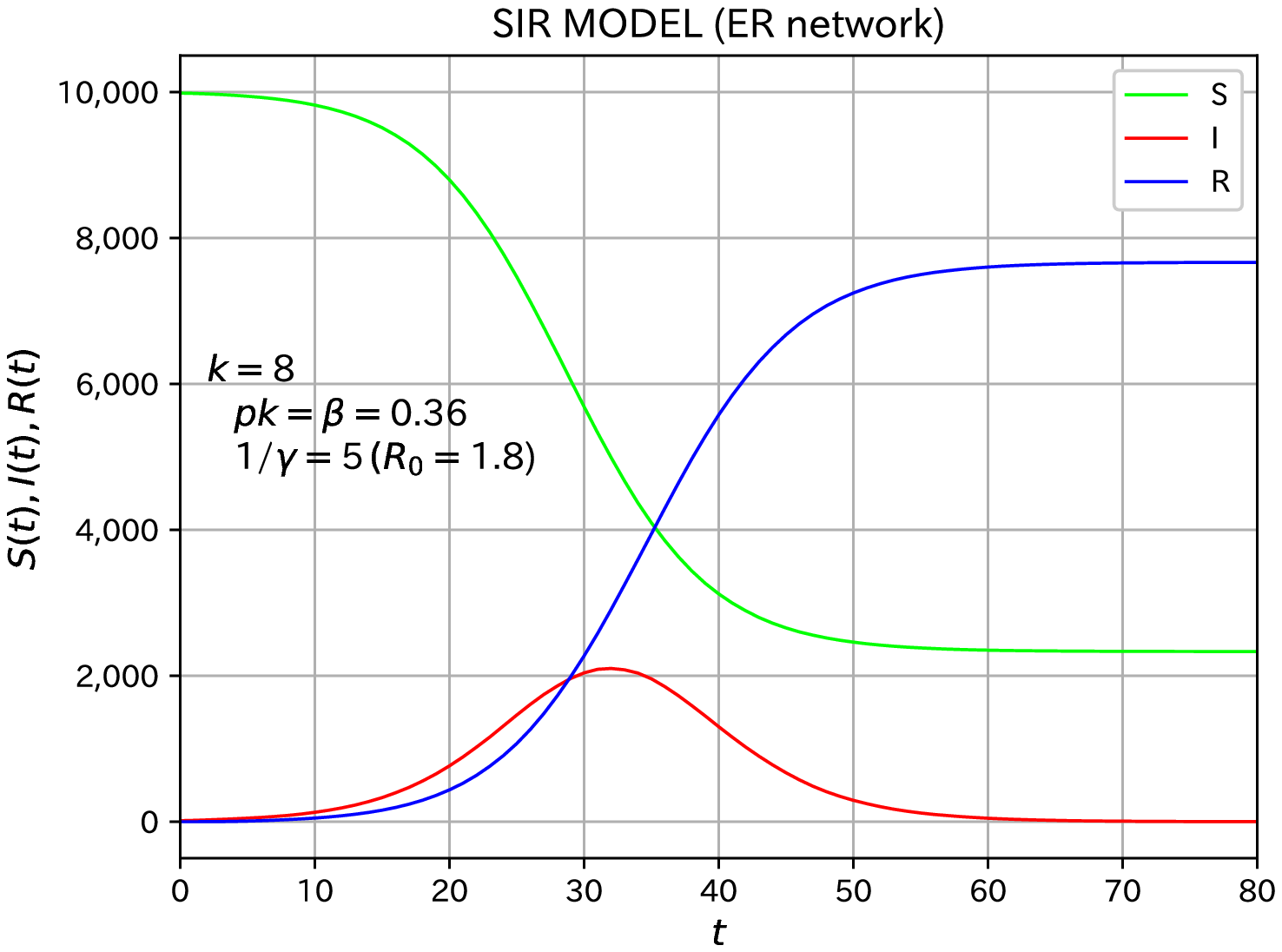}
\caption{
The simulational results of the microscopic SIR model 
on the ER network (reference system with no variants).  
(a) the plot of all 100 samples assumed to be infected with 
the virus with $R_0=1.8$.
(b) the average over 100 samples. 
Initially, 10 individuals were set to be infected. 
}
\label{fig:ER_var0}
\end{figure}

As a reference system, we deal with the case where there are no variants. 
The initial condition ($t=0$) is that 10 randomly selected individuals 
are infected. The time evolution of the number of individuals of 
the three types (S, I, and R) is shown in Figure \ref{fig:ER_var0}. 
We performed simulations for 100 samples, and the time evolution 
of all samples are plotted in Figure \ref{fig:ER_var0}a.
We observe a variation in the time evolution for each sample. 
On the other hand, Figure \ref{fig:ER_var0}b is an average plot 
of 100 samples.
The number of infected individuals (I) increases with time, 
reaches a peak, and gradually decreases, which is the same behavior 
as that for the SIR model of the differential equation. 
The microscopic SIR model on a random network is considered to 
reproduce the macroscopic SIR model almost quantitatively. 
The final value of the total number of infected individuals 
($R(\infty)$) is 0.732 from Equation (\ref{final}), 
the final size equation of the SIR model \cite{Metz86}, 
in the case of $R_0=\beta/\gamma=1.8$, 
and the measured value 
for the present microscopic model 
on the ER network is about 0.77. 
As a related argument, the final size was discussed 
in the framework of a stochastic epidemic model~\cite{Britton2010}. 
In the Reed-Frost model~\cite{Schwabe,Abbey}, which is 
a typical example of a stochastic epidemic model, 
the final size can be treated analytically. 
Britton {\it et al.} discussed the final size of 
stochastic epidemic models on networks \cite{Britton2007}. 
In the previous study \cite{Okabe2021}, for a small number 
of initially infected individuals, 
some examples showed the behavior such that the infection 
vanishes quickly and does not spread throughout the network. 
This behavior is regarded as the absorbing state \cite{Marro,Henkel,Mata} 
in the contact process \cite{Harris}. 
We chose 10 initially infected cases to avoid the situation 
of the absorbing state. 

\begin{figure}[h]
\centering
{\bf a} \hspace*{7cm} {\bf b}\hspace*{7cm}

\includegraphics[width=7cm]{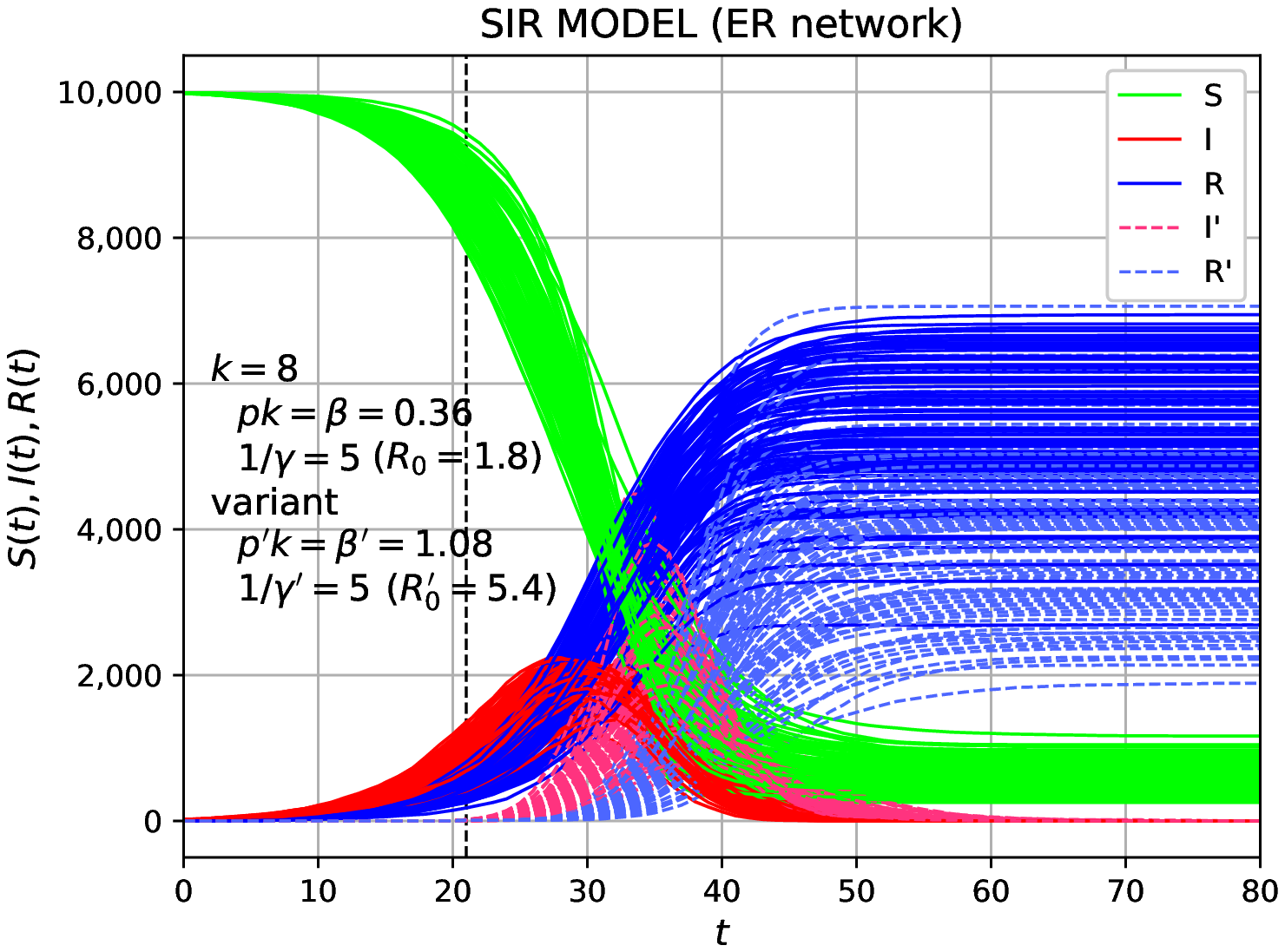}
\includegraphics[width=7cm]{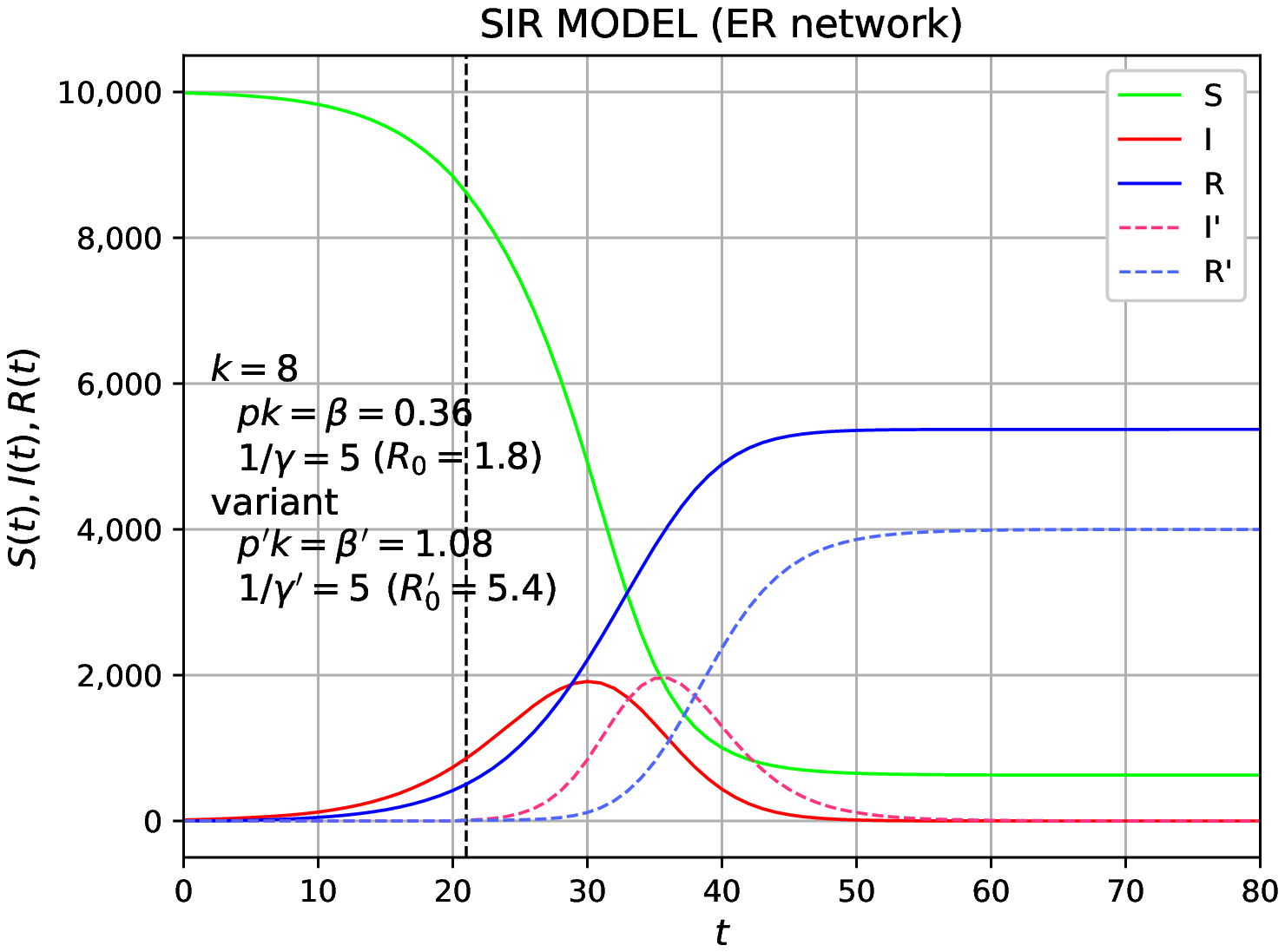}
\caption{
Effects of variants for the spread of the epidemic disease 
on the ER network. 
Suppose that 10 individuals are infected with the variant ($R_0'=5.4$) 
by an external factor at $t=21$.
(a) the plot of all 100 samples.
(b) the average of the 100 samples.
The red and blue dashed lines indicate the values of the variant.}
\label{fig:ER_var1}
\end{figure}

We next consider the effects of variants. Suppose that 10 susceptible 
individuals are infected with the variant due to external factors 
at $t=21$. We choose 10 individuals randomly among the non-infected. 
The variant is assumed to be 3.0 times more infectious 
with $\beta'=1.08$, and the average infection period is chosen as 
the same value as that of the original virus, $1/\gamma'=5.0$. 
Then, the basic reproduction number of the variant becomes $R_0'=5.4$.
The situation at $t=21$ is that about 900 individuals are infected, 
500 individuals are recovered, and 8600 individuals are not infected.
In Figure \ref{fig:ER_var1}, the number of individuals infected 
with the variant (I') 
and those who recovered from the variant (R') are shown 
in the dashed line. The time when the variant is added 
is indicated by the vertical black dashed line.
The variant infection starts to spread at $t=21$, but as can be seen 
in Figure \ref{fig:ER_var1}a, which plots all 100 samples, 
there is a large sample dependence. 
Figure \ref{fig:ER_var1}b, which is the average of 100 samples, 
shows the general trend.
The value of $R(\infty)$ represents the final total number of 
infected individuals of the original virus, 
and the value of $R'(\infty)$ represents the total number 
of individuals infected with the variant.
The value of $R(\infty)$ slightly decreases compared to the case 
without the variant. 
Compare the solid blue line in Figure \ref{fig:ER_var1}b 
with that in Figure \ref{fig:ER_var0}b. 

\begin{figure}[h]
\centering
\includegraphics[width=7cm]{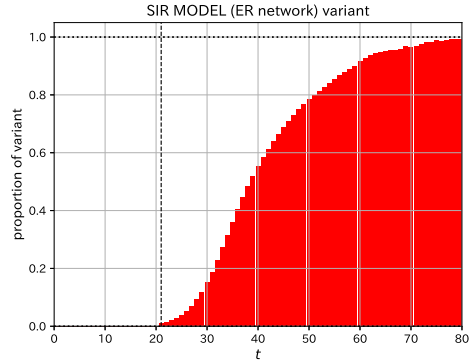}
\caption{
Time variation of the proportion of variants among the infected.
The data shown in Figure \ref{fig:ER_var1} was used 
to make this graph.
}
\label{fig:ER_prop}
\end{figure}

An increase and decrease in the proportion of variants 
are often discussed. 
The calculated time variation of the proportion of variants 
from the data in Figure \ref{fig:ER_var1} is plotted 
in Figure \ref{fig:ER_prop}.
An increase in the proportion of variants is observed, 
which means that the infection of variants has spread. 

\begin{figure}[h]
\centering
{\bf a} \hspace*{7cm} {\bf b}\hspace*{7cm}

\includegraphics[width=7cm]{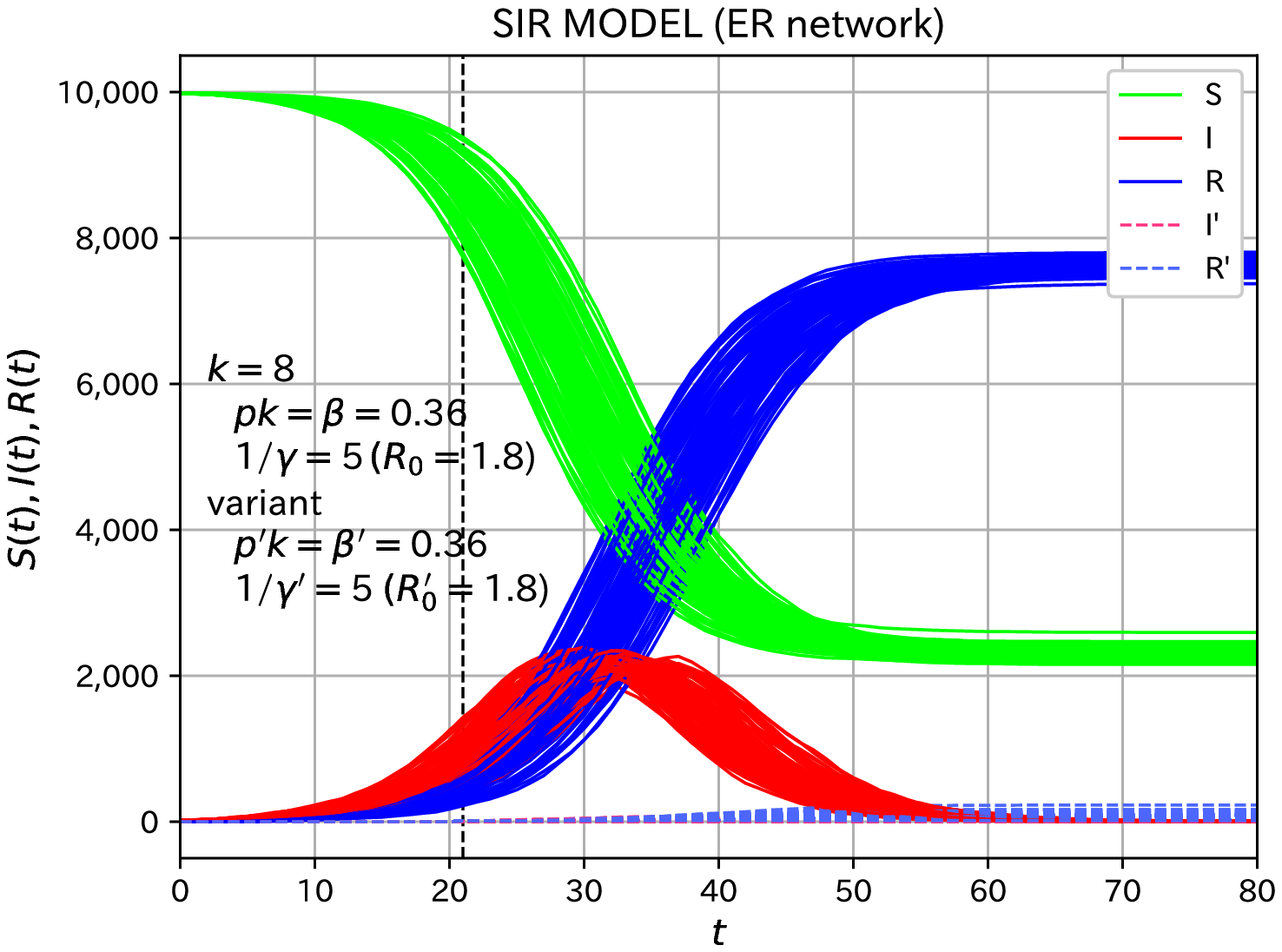}
\includegraphics[width=7cm]{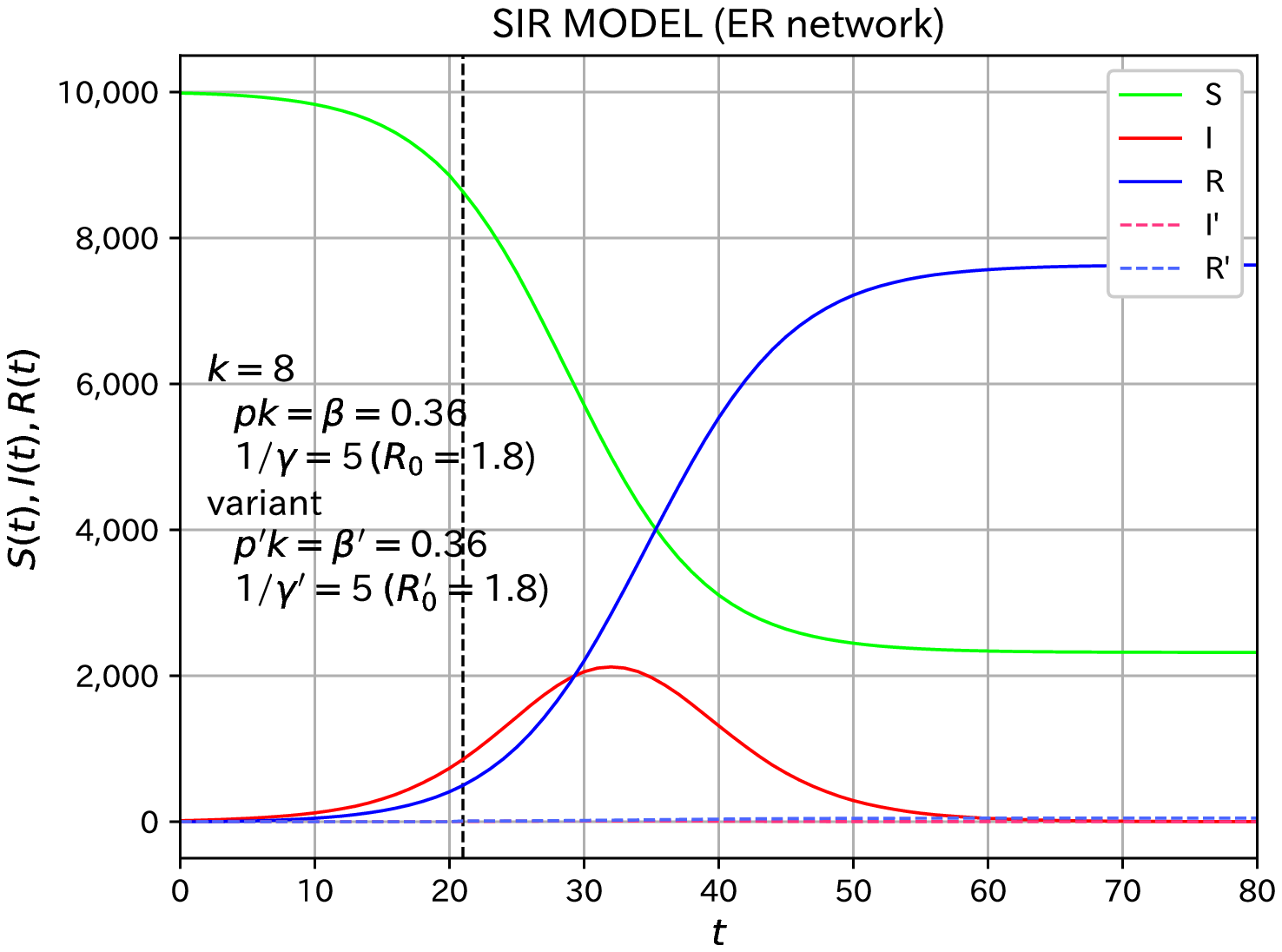}
\caption{
Effects of variants for the spread of the epidemic disease 
on the ER network. 
Suppose that 10 individuals are infected with the original virus ($R_0'=1.8$) 
by an external factor at $t=21$.
(a) the plot of all 100 samples.
(b) the average of the 100 samples.
The red and blue dashed lines are additional values.}
\label{fig:ER_var2}
\end{figure}

For comparison, let us consider the case where the infectivity 
of the variant to be added is the same as that of the original species.
Suppose that at $t=21$, 10 people are infected by the variant 
with $R_0'=1.8$, the same as the original species.
Figure \ref{fig:ER_var2} shows the time evolution of 
the infected individuals.
As can be seen from Figure \ref{fig:ER_var2}, in this case, 
the variant infection does not spread for all 100 samples.
The spread of the additional variant is regarded as in the absorbing state. 
The addition of 10 infected individuals with the variant 
to 900 infected individuals will have little effect.

We have shown that when the infectivity of the variant is strong, 
even if a small number of individuals are infected, 
it leads to the spread of the infection. 
On the other hand, when the infectivity of the variant is comparable 
to that of an original virus, it does not spread, 
and this is a dynamic effect. 
So what happens when the infectivity is intermediate$?$ 
The results of measurements for intermediate values 
of the basic reproduction number of variants are shown 
in the supplementary information. 
There, the time evolution graphs for the variants with 
$R_0'=4.95$, $R_0'=4.5$, $R_0'=4.05$, $R_0'=3.6$, $R_0'=3.15$, 
$R_0'=2.7$, and $R_0'=2.25$ are also given.

\begin{figure}[h]
\centering
\includegraphics[width=7cm]{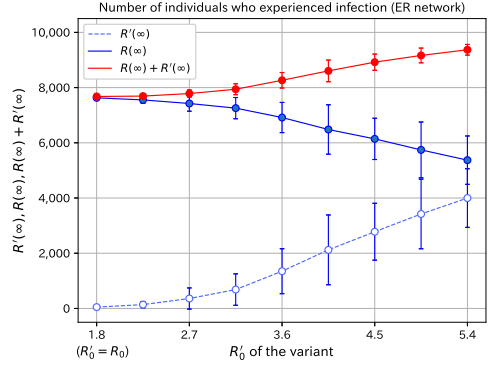}
\caption{
Dependence of the final number of infected individuals on $R_0'$ 
of the variant.  The infectivity of the variant is the same as 
that of the original virus for $R_0'=1.8$.}
\label{fig:ER_depend}
\end{figure}

We examine the behavior of the spread of variants systematically.
The dependence of the final number of infected 
on the basic reproduction number $R_0'$
of the variant is plotted in Figure \ref{fig:ER_depend}. 
The number of infected individuals of the variant is shown 
by the dashed blue line and the number of infected individuals 
of the original species by the solid blue line. The standard 
deviation of the 100 samples is indicated by the error bars.
The infectivity of the variant is the same as 
that of the original virus for $R_0'=1.8$.
It can be seen that the number of infected individuals of the variant 
increases with the infectivity 
and the fluctuation also increases. 
The increase is not linear; that is, when the infectivity 
becomes slightly larger, the spread of infection is small, 
but as the infectivity further increases, the increase is non-linear.
On the other hand, the number of the infected individuals 
with the original species 
decreases slightly. 
This is because individuals who would have been infected 
with the original species if there had been no new infections 
of the variant will be infected with the variant. 
The sum of the infected individuals with the variant and 
those with the original species is shown by the red line, 
and the sum increases. However, the fluctuation of the sum 
is not necessarily large.

To summarize what we have shown in this subsection, we can say the following. 
When the number of infected individuals reaches about 900 
(the total number is 10000), a small number of individuals
infected with the variant are added. 
If the variant is highly infectious, it will spread 
to the wide space from the 10 additional infected individuals.
By systematically examining the variants whose basic reproduction 
numbers are 1.0, 1.25, 1.5, 1.75, 2.0, 2.25, 2.5, 2.75, and 3.0 times 
that of the original species, 
we showed that the infection increases non-linearly 
with the basic reproduction number.
The lack of infection to spread in the case of variants 
whose infectivity is not much different from that 
of the original species is due to the dynamic effect of infection, 
that is, the absorbing state.

\subsection{Simulation of the microscopic model 
for the variants on the BA network}

Next, we consider the case of the BA network, a scale-free network. 
We choose the network which is similar to the case of the ER network. 
The total number of nodes (individuals) is $N=10000$ and 
the average number of degrees is $\l k \r=8$. 
The spread of infection is more rapid in the case of 
scale-free networks because of a hub structure \cite{Okabe2021}. 
Thus, for the probability of infection $p$, we choose a value 
smaller than that for the ER network, that is, $p=1/25$.
This leads to $\beta = \l k \r p = 0.32$ in terms of 
a rate constant of the SIR model. 
The average infected period, $1/\gamma$, is again chosen 
as 5.0 days. 
Then, the basic reproduction number becomes $R_0=\beta/\gamma=1.6$.

\begin{figure}[h]
\centering
{\bf a} \hspace*{7cm} {\bf b}\hspace*{7cm}

\includegraphics[width=7cm]{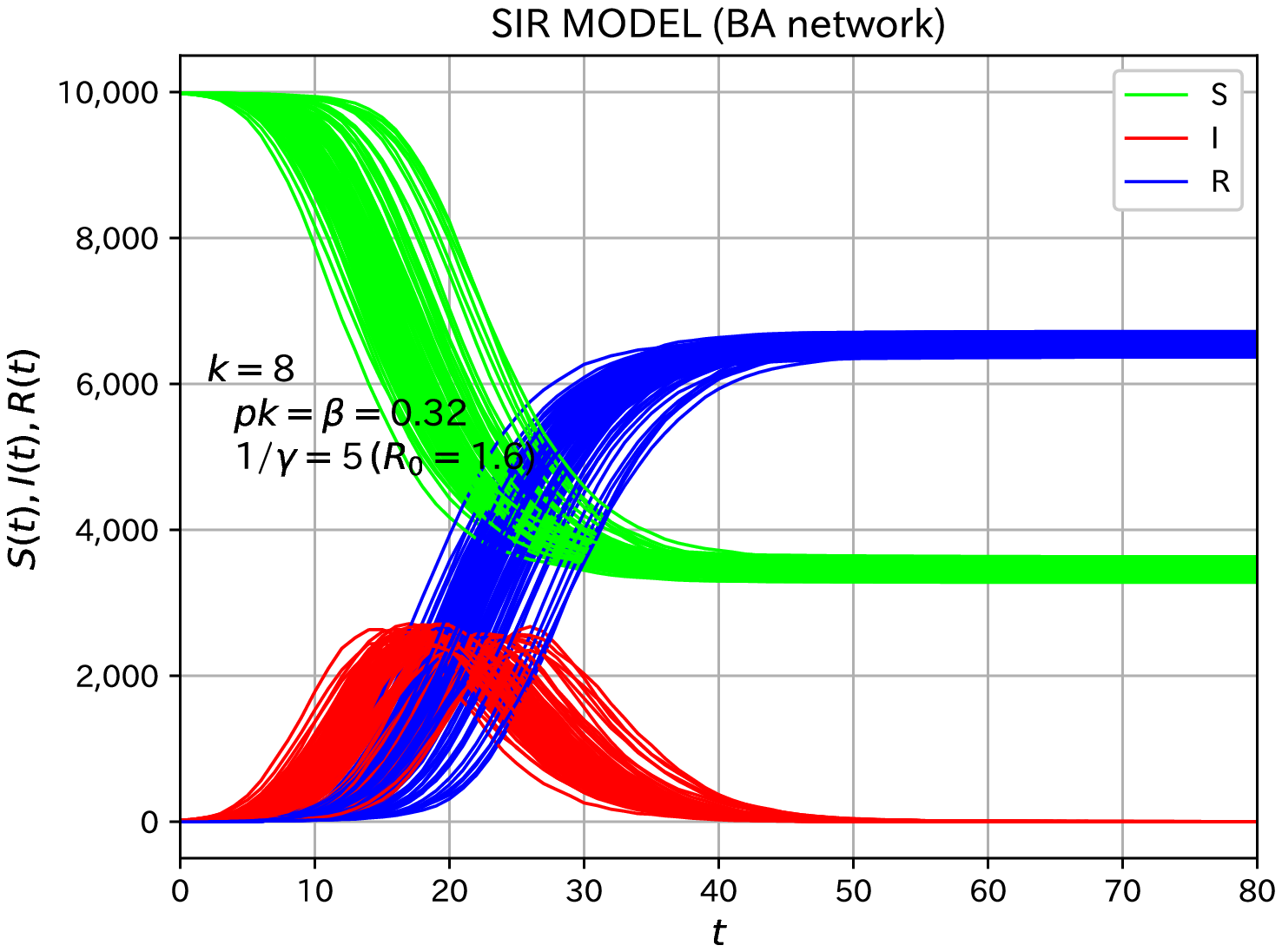}
\includegraphics[width=7cm]{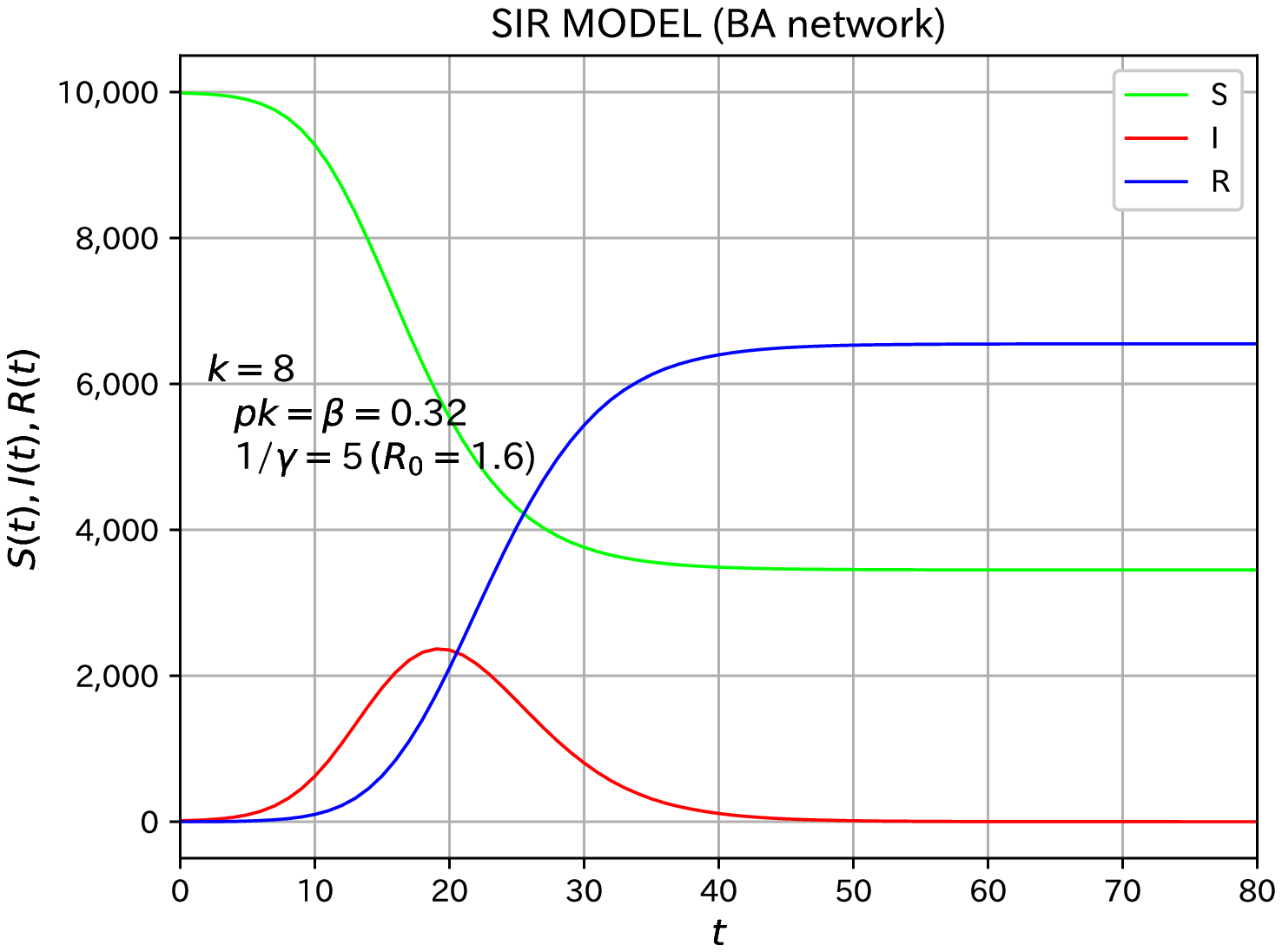}
\caption{The simulational results of the microscopic SIR model 
on the BA network (reference system with no variants).  
(a) the plot of all 100 samples assumed to be infected with 
the virus with $R_0=1.6$.
(b) the average over 100 samples. 
Initially, 10 individuals were set to be infected. }
\label{fig:BA_var0}
\end{figure}

We treat the case where there are no variants as a reference system. 
The conditions are the same as the case of the ER network. 
The time evolution of the number of individuals of 
the three types (S, I, and R) is shown in Figure \ref{fig:BA_var0}. 
We performed simulations for 100 samples. 
We plot the time evolution of all 100 samples 
in Figure \ref{fig:BA_var0}a, 
whereas the average of 100 samples is plotted 
in Figure \ref{fig:BA_var0}b.

\begin{figure}[h]
\centering
{\bf a} \hspace*{7cm} {\bf b}\hspace*{7cm}

\includegraphics[width=7cm]{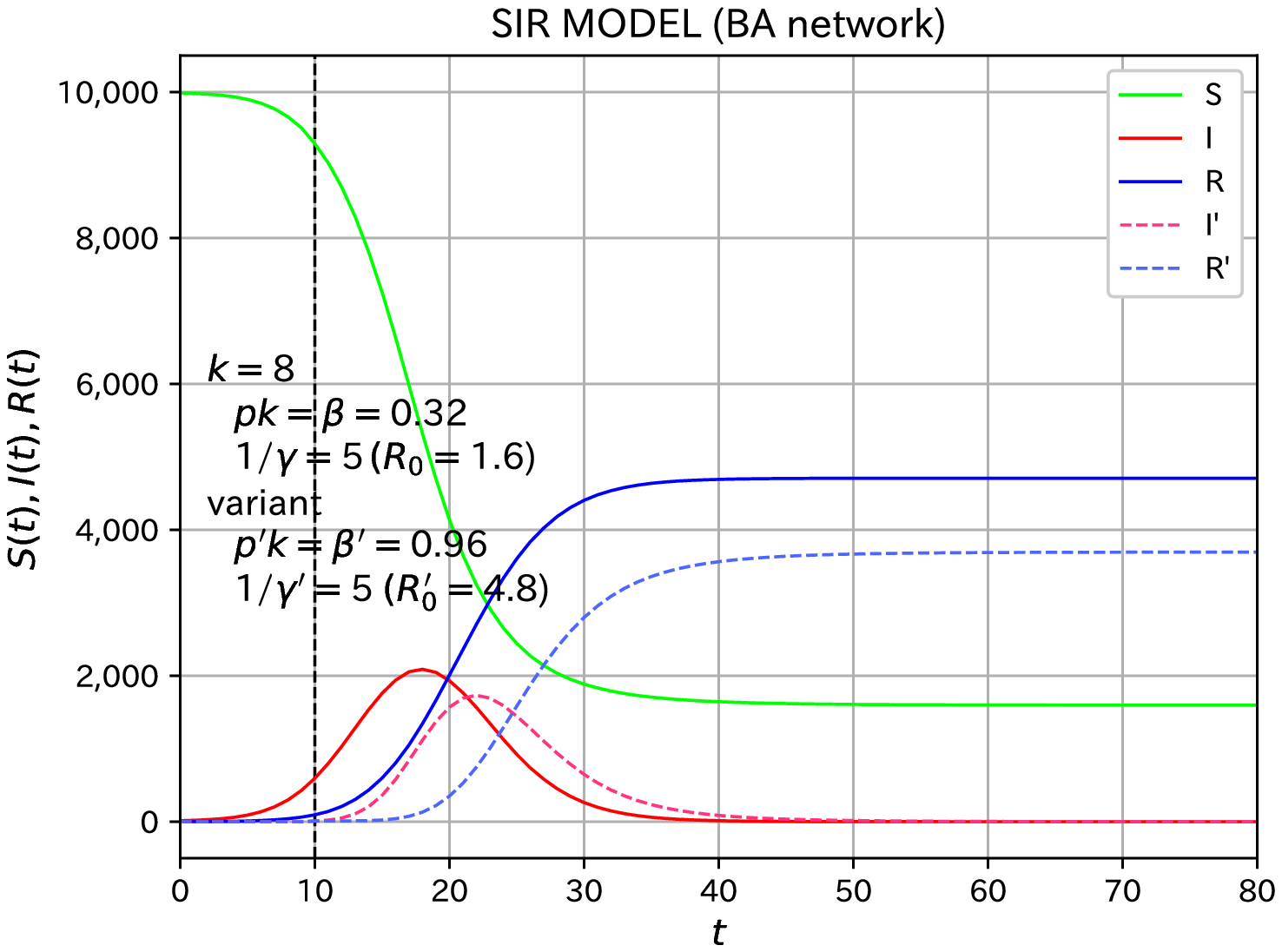}
\includegraphics[width=7cm]{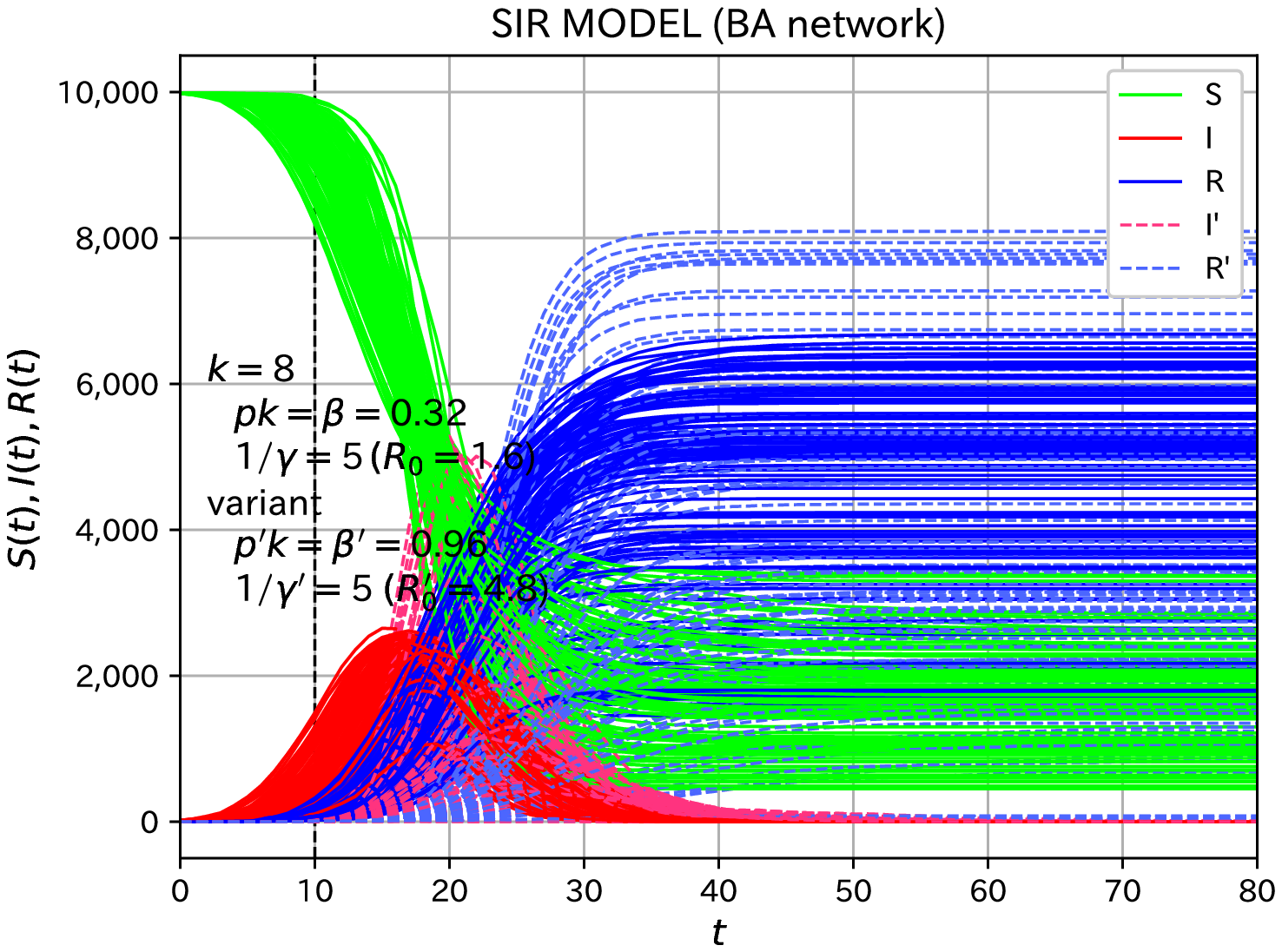}
\caption{
Effects of variants for the spread of the epidemic disease 
on the BA network. 
Suppose that 10 individuals are infected with the variant ($R_0'=4.8$) 
by an external factor at $t=10$.
(a) the plot of all 100 samples.
(b) the average of the 100 samples.
The red and blue dashed lines indicate the values of variant.}
\label{fig:BA_var1}
\end{figure}

We turn to the investigation of the effects of variants. 
Suppose that 10 susceptible individuals are infected 
with the variant due to external factors at $t=10$. 
The variant is assumed to be 3.0 times more infectious 
with $\beta'=0.96$. 
The average infection period is chosen as $1/\gamma'=5.0$, 
which leads to $R_0'=4.8$. 
The situation at $t=10$ is that about 620 individuals are infected, 
100 individuals are recovered, and 9280 individuals are not infected.
This is similar to the case of ER network at $t=21$ shown 
in Figure \ref{fig:ER_var1}.
We plot the time variation of the spread of infection on the BA network 
in Figure \ref{fig:BA_var1}. The number of individuals infected 
with the variant (I') and those who recovered from the variant (R') 
are shown in the dashed line. 
The overall behavior of the BA network is similar to the case 
of the ER network.

\begin{figure}[h]
\centering
\includegraphics[width=7cm]{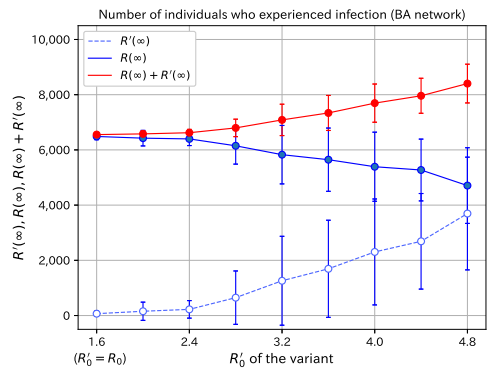}
\caption{
Dependence of the final number of infected individuals on $R_0'$ 
of the variant. The infectivity of the variant is the same as 
that of the original virus for $R_0'=1.6$.}
\label{fig:BA_depend}
\end{figure}

The results of measurements for various $R_0'$ of the variant 
on the BA network are shown 
in the supplementary information. 
There, the time evolution graphs for the variants with 
$R_0'=4.4$, 4.0, 3.6, 3.2, 2.8, 2.4, 2.0, and 1.6 are also given.
The basic reproduction number $R_0'=1.6$ is the same number 
as the original virus. 
Summarizing the results for various $R_0'$, we plot the dependence of 
the final number of infected on the basic reproduction number $R_0'$
of the variant in Figure \ref{fig:BA_depend}. 
The number of infected individuals of the variant is shown 
by the dashed blue line, the number of infected individuals 
of the original species by the solid blue line, 
and the sum of the two by the red line. The standard deviation 
of the 100 samples is indicated by the error bars.
The infectivity of the variant is the same as 
that of the original virus for $R_0'=1.6$.
From Figure \ref{fig:BA_depend}, we observe that if the variant 
is highly infectious, the number of individuals infected 
with the variant increases. From the systematical study 
with infectivity of 1.0, 1.25, 1.5, 1.75, 2.0, 2.25, 2.5, 2.75, and 3.0 times, 
we see that the effect is not simply proportional 
to the infectivity. 
We also observe that the number of individuals 
infected with the original species decreases slightly 
and the total number of the infected increases. 
The fluctuation in the number of infected individuals 
becomes significant when the infectivity of the variant is strong. 
However, the fluctuation of the sum of the variant and 
the original species is not necessarily significant. 
The effect of a variant on the BA network is essentially the same 
as that of a variant on the ER network. The characteristics of 
the scale-free network are not particularly evident.

\section{Discussion}

We studied the spread of variants of the virus by performing 
numerical simulations of the microscopic model on the network. 
In the middle of a simulation of infectious disease 
on the network, we added a variant that is 
more transmissible than the original lineage. 
When a highly infectious variant is added, the variant spreads quickly. 
It is noteworthy that the rate of spread is not linear in the 
infectivity of the variant but it is non-linear. 
If a variant is slightly more transmissible than 
the original lineage, the virus does not spread widely. 
It is related to a non-equilibrium 
phase transition of the epidemic 
dynamics on the complex network between a disease-free 
(absorbing) state and an active stationary phase \cite{Marro,Henkel,Mata}. 
This cannot be accounted for by the compartmental model 
of epidemiology. 
We performed simulations both on the ER model, a random network, 
and the BA model, a scale-free network. 
It was shown that the existence of the hub for a scale-free network 
stimulates the rapid increase of the infection \cite{Okabe2021}. 
But the effect of the variants for the BA network is essentially 
similar to the case of the ER network. 
It should also be noted that there is a lot of fluctuation 
in the spread of the more infectious variants. 
The greater fluctuation is due to the network structure.
In summary, the reason why the more infectious variants spread, 
while the less infectious variants do not, has been clarified.

Although the present work is for a model system 
with some parameters, we may learn some lessons 
for a real-world strategy to fight against the pandemic. 
The spread of a slightly more infectious variant is not significant. 
Variants that are more infectious than the original lineage 
will spread the infection in a non-linear fashion. 
Of course, whether or not the variant is severe 
is an important factor. 
It can be said that the major role of overseas quarantine is 
to identify variants. For highly infectious variants, 
it is necessary to identify them by genetic testing 
and to isolate infected individuals. 
To conduct special contact tracing is of particular importance
in the case that they spread throughout the city.
Those variants must be monitored more carefully. 
Measures need to be taken to control the spread of the virus and its variants. 
Travel restrictions can be effective in reducing the spread of the virus. 
Contact with people over long distances is associated with 
the presence of hubs in the network. 
As shown in the previous paper \cite{Okabe2021}, isolation of hub sites 
can lead to transmission control. This is also the case for variants, 
for which travel restrictions are effective.
However, to ultimately control the epidemic, it may be necessary 
to greatly accelerate vaccine roll-out.

\section{Methods}

\subsection{Differential equation of SIR model}
\label{sub:SIR}

We perform simulation of the microscopic model of epidemic disease, 
which corresponds to the compartmental model of 
macroscopic variables. We consider SIR model 
\cite{Kermack,Bailey,Diekmann2000,Okabe,Okabe2021} 
as a compartmental model. 

We consider a closed society of $N$ individuals 
and classify individuals into three types: 
susceptible (S), infected (I), and recovered (R).
Infected individuals can only transmit the virus to 
susceptible individuals. Once infected individuals have recovered 
(or have passed away), they can no longer infect others 
and cannot be reinfected. 
Let $S(t)$, $I(t)$, and $R(t)$ denote the number of individuals 
in three states S, I, and R as a function of time $t$. 
In the SIR model, the time evolution is described 
by the following simultaneous differential equations. 
\begin{eqnarray}
 \frac{dS(t)}{dt} &=& - \frac{\beta}{N} S(t)I(t),
 \label{eq1} \\
 \frac{dI(t)}{dt} &=& \frac{\beta}{N} S(t)I(t)- \gamma I(t),
 \label{eq2} \\
 \frac{dR(t)}{dt} &=&  \gamma I(t),
 \label{eq3} 
\end{eqnarray}
where $\beta$ is the rate of infection and $\gamma$ the rate of 
recovery (the rate of quarantine). 
We consider that each person is in contact with $k$ persons 
per unit time (day), and the probability of infection for each contact 
is set as $p$. Then, $\beta$ is given as $kp$. 
We assume that the total number of individuals is set to be constant 
such that 
\begin{equation}
  S(t)+I(t)+R(t)=N. 
\label{eq:const}
\end{equation}
Here, we do not consider the birth and death processes. 

The SIR equation, (\ref{eq1})$\sim$(\ref{eq3}), 
is said to have been given an exact solution by Harko {\it et al.} 
in 2014~\cite{Harko}.  
However, from the standpoint of differential equation theory, 
it was known to be solvable before that~\cite{Hirsch}.
From Equations~(\ref{eq1}) and (\ref{eq2}), we obtain
\begin{equation}
  \frac{dI}{dS} = \frac{- (\beta/N) S + \gamma}{(\beta/N) S}
                = -1 + \frac{\gamma}{\beta} \Big( \frac{N}{S} \Big).
\label{dydx}
\end{equation}
The integration with respect to $S$ yields 
\begin{equation}
  I = - S + \frac{\gamma}{\beta} N \ln S + C,
\label{y_vs_x}
\end{equation}
where $C$ is an integral constant. 
If we insert Equation (\ref{y_vs_x}) into Equation (\ref{eq1}), 
we obtain a closed expression for $S(t)$.
We note that the ratio of $\beta$ and $\gamma$ appears for rate constants. 
The number $R_0$ defined by
\begin{equation}
   R_0 = \frac{\beta}{\gamma}
\label{basic_repro}
\end{equation}
is known as the basic reproduction number 
\cite{Diekmann90,Diekmann2000,Dietz93}, 
and the number of infected individuals increases when $R_0 > 1$, 
whereas it decreases when $R_0 < 1$.

It is noteworthy that $I(t)$ is the number of infected individuals, not the number 
of new infected individuals announced every day. 
It should be noted that the number of new infections can be discussed 
in the framework of the SIR model \cite{Okabe}.

We here make a comment on the final size equation. 
We have the relation for the final value of $R(\infty)$. 
In the limit of $N_1 \to N$, we obtain
\begin{equation}
  1 - R(\infty)/N = \exp \Big[- \frac{\beta}{\gamma}(R(\infty)/N) \Big].
\label{final}
\end{equation}
This relation is known as the final size equation \cite{Metz86}.

\subsection{Method of simulation of the microscopic SIR model}
\label{sub:micro}

Let us consider a simulation of microscopic infectious disease propagation 
in a network, in which each node assumes the state of infectious disease 
and changes its state stochastically. 
The rules for updating the state correspond to the macroscopic SIR model 
given by the differential equations, 
and the parameters are chosen to be equivalent 
so that the macroscopic and microscopic SIR models can be compared. 
The actual procedure, which is similar to that 
by Herrmann and Schwartz~\cite{Herrmann}, is as follows \cite{Okabe2021}:

\begin{enumerate}
\item
Generate a network.
\item
At $t=0$, an individual or individuals are initially infected (I). 
\item
A susceptible individual (S) will be infected (I) 
with a probability $p$ if a connecting individual (one of $k$) 
is infected (I). 
We do this for all nodes (non-infected) and for all connections 
($k$) of each node. 
In terms of the SIR model, the parameter $\beta$ 
is $\beta=kp$. 
\item
An infected individual (I) will recover (R) in $1/\gamma$ days on average. 
The infected period is chosen by a Poisson distribution 
with an average of $1/\gamma$. 
\item
At each time $t$, the processes 3, 4 are repeated.
\item
The time sequence obtained from the above procedure is 
regarded as a single sample.  Simulations are performed 
for several samples. 
\end{enumerate}

In the simulation to study the effect of the variant, 
we will assume that some individuals are infected by the variant 
at time $t=t_{\rm var}$. 
We will denote by I' those who are infected by the variant 
and by R' those who recover from the variant.
Modify the processes 2-4 of the simulation 
as follows:
\begin{enumerate}
\setlength{\leftskip}{1.5mm}
\renewcommand{\labelenumi}{\arabic{enumi}'.}
\setcounter{enumi}{1}
\item
At $t=t_{\rm var}$, an individual or individuals are infected with the variant (I'). 
\item
A susceptible individual (S) will be infected by the original virus (I) 
with a probability $p$ if a connecting individual 
is infected by the original virus (I).  If a connecting individual 
is infected by the variant (I'), a susceptible individual (S) will be 
infected by the variant (I') with a probability $p'.$ 
\item
An infected individual by the original virus (I) will be 
recovered (R) in $1/\gamma$ days on average. 
The infected period is chosen by a Poisson distribution 
with an average of $1/\gamma$. 
An infected individual by the variant of the virus (I') will be 
recovered (R') in $1/\gamma'$ days on average. 
\end{enumerate}

Then, the condition in Equation (\ref{eq:const}) that the total number 
of individuals of various types is conserved is modified as
\begin{equation}
  S(t)+I(t)+R(t)+I'(t)+R'(t)=N. 
\label{eq:const2}
\end{equation}
We perform the simulation of the microscopic model 
using the procedures described above.
We here make a short comment on the distribution of the infection period. 
The geometric distribution for a discrete-time model was used to 
evaluate the impact of waiting-time distributions on epidemiological 
processes~\cite{Hernandez-Ceron}. 
On the other hand, Herrmann and Schwartz~\cite{Herrmann} used a distribution referring to the covid-19 data to perform microscopic simulations 
on complex networks. 
In the present paper, we employed the Poisson distribution 
as a model, which does not have a very long time tail. 
It is the effect of the variants that we are discussing.

\section*{Acknowledgements}
The authors thank Hiroyuki Mori for valuable discussions.

\section*{Author contributions}
These authors contributed equally to this work. 
All authors approved the final manuscript.

\section*{Corresponding author}
Correspondence to Yutaka Okabe.

\section*{Competing interests}
The authors declare no competing interests.

\section*{Additional information}
{\bf Supplementary information} is available for this paper at ***.

\end{document}